\relax
\documentclass[letterpaper]{article} % DO NOT CHANGE THIS
\usepackage{aaai21}  % DO NOT CHANGE THIS
\usepackage{times}  % DO NOT CHANGE THIS
\usepackage{helvet} % DO NOT CHANGE THIS
\usepackage{courier}  % DO NOT CHANGE THIS
\usepackage[hyphens]{url}  % DO NOT CHANGE THIS
\usepackage{graphicx} % DO NOT CHANGE THIS
\usepackage{subfigure}

\usepackage{natbib}  % DO NOT CHANGE THIS AND DO NOT ADD ANY OPTIONS TO IT
\usepackage{caption} % DO NOT CHANGE THIS AND DO NOT ADD ANY OPTIONS TO IT
\title{What Kind of Person Wins the Turing Award?}
\author{Anonymous submission}
\author{
    Zhongkai Shangguan\textsuperscript{\rm 1*}, Zihe Zheng\textsuperscript{\rm 2}\thanks{The two authors contributed equally.}, Jiebo Luo\textsuperscript{\rm 3}
    \\
}
\affiliations{
    \textsuperscript{\rm 1}Department of Electrical and Computer Engineering, \textsuperscript{\rm 2} Goergen Institute for Data Science, \textsuperscript{\rm 3} Department of Computer  Science\\
    \textsuperscript{\rm 1}\textsuperscript{\rm 2}\textsuperscript{\rm 3}University of Rochester, Rochester, NY, 14627\\
    \textsuperscript{\rm 1} zshangg2@ur.rochester.edu, \textsuperscript{\rm 2} zzheng18@u.rochester.edu, \textsuperscript{\rm 3} jluo@cs.rochester.edu\\
}

\begin{document}

\maketitle

\begin{abstract}
Computer science has grown rapidly since its inception in the 1950s and the pioneers in the field are celebrated annually by the A.M. Turing Award. In this paper, we attempt to shed light on the path to influential computer scientists by examining the characteristics of the 72 Turing Award laureates. To achieve this goal, we build a comprehensive dataset of the Turing Award laureates and analyze their characteristics, including their personal information, family background, academic background,  and industry experience. The FP-Growth algorithm is used for frequent feature mining. Logistic regression plot, pie chart, word cloud and map are generated accordingly for each of the interesting features to uncover insights regarding personal factors that drive influential work in the field of computer science. In particular, we show that the Turing Award laureates are most commonly white, male, married, United States citizen, and received a PhD degree. Our results also show that the age at which the laureate won the award increases over the years; most of the Turing Award laureates did not major in computer science; birth order is strongly related to the winners’ success; and the number of citations is not as important as one would expect. %Our contributions include: we construct a diverse dataset of the Turing Award laureates and provide general as well as detailed analysis of their features.
\end{abstract}

\section{Introduction}
The field of Computer Science is one of the fastest growing disciplines in the world. As a very important part of the Digital Revolution \cite{digitalrev}, the development of computer science is driven by scientific questions, technological innovation and societal demands. It has direct impact on people's lives and the potential to fundamentally change the traditional way of life. 
As we live in a digital age, Computer Science is permeating into various fields of social sciences and natural sciences, leading to new changes in social production, creating new spaces for human life, bringing new challenges to national governance, and profoundly changing the global industry, economy, interests, and security patterns. 
Who are those people that have significant impact on the design and direction of computer technology that we use? Can we find some rules or best practices to promote further development of computer science? These questions prompt people to study from outside computer science itself, e.g., from the perspective of history and the sociology of science \cite{akmut}. In this study, we are analyzing from the perspective of data science, more specifically, we try to discover patterns from the diverse information of Turing Award Laureates.

Turing Award, in full A.M. Turing Award, is Association for Computing Machinery (ACM)’s oldest and most prestigious award. The award is named after Alan M. Turing, a British pioneer of computer science, mathematics and cryptography \cite{alanturing}, and is presented annually to individuals who have made lasting contributions of a technical nature to the computing community \cite{TuringOfficial}. The Turing Award is recognized as the highest distinction in computer science, and is referred to as the computer science equivalent of the Nobel Prize \cite{xia2020turing}. The selection process for the Turing Award is very strict. Although considering the long-term impact of the nominee’s work, there should be a particularly outstanding and trend-setting technical achievement, which is the main proposition of the award. The successful nomination of the Turing Award includes substantive support letters from prominent individuals in the field or related fields representing a wide range of candidates. These letters provide clear evidence of the candidate's lasting influence \cite{TuringNominate}.

Alan Jay Perlis was the first prize winner, who contributed significantly to the fact that computer science was taught as an independent subject at American universities. The youngest winner was Donald Knuth, who convinced the jury with “Computer Programming as an Art” and won Turing Award in 1974 at the age of 36. In 2006, Turing Award was first given to a woman, Frances Elisabeth Allen, for her work on the theory and practice of compiler optimization \cite{t2informatic}. As of 2020, the Turing Award has been presented 54 times since 1966, with 72 Turing Award winners. In our study, we aim to provide insights on the key personal factors of winning a Turing Award by investigating and analyzing the background of the past winners.

This study collects the relevant information of the Turing Award winners over the years (until 2019) in order to have a comprehensive understanding of the winners of the Turing Award. We conduct a preliminary analysis of the winners’ bibliography, family information, academic background, personal experience, and so on, hoping to provide reference of the discipline development and training direction in the field of computer science.
The rest of the paper is organized as follows. \textbf{Background and Related Work} introduces background of the Turing Award. The dataset and methodology are presented in detail in \textbf{Methods}. We show our fine-grained analysis in the \textbf{Results} section. Finally, we  conclude our work and discusses future directions.

The contributions of this work are summarized as follows:
\begin{enumerate}
    \item We construct the latest dataset that contains diverse data of all Turing Award Laureates (as of 2019) using a number of web resources.
    \item We provide an in-depth analysis of the characteristics of the Turing Award Laureates and uncover interesting insights regarding personal factors that drive influential work in the field of computer science.
\end{enumerate}

\section{Background and Related Work}
Scientists in various fields have made many attempts to find the common features of high achieving scientists and how they developed those desirable characteristics. Most of them are from sociological perspective, focusing on family background \cite{family}, academic network of scientists \cite{wagner}, the years in which training was received \cite{golden}, and affiliated institutions \cite{schlag}. Many of these work focus on Nobel Prize winners, yet they still inspire our research and help us with feature selection in our research. \citet{nobel2016} studied the educational background of the Nobel Prize winners and found that undergraduates from small and elite institutes are the most likely to win a Nobel Prize. \citet{clark} studied on Nobel Prize winners and found that eminent scientists tend to be earlier born when family size is not controlled while in smaller families the later born are more likely to be eminent in science. A sociographic study \cite{berry} explored the cultural context in which the stable and introvert personality of most Nobel Prize winners grew: they found that scientific theories of achievement and national planning in science should be accountable. The best nations and institutions for revolutionary science was studied by \cite{charlton}: the dataset includes Nobel Prizes, Fields Medals, Lasker Awards and Turing Awards and they found that MIT, Stanford and Princeton are the top institutions and that the USA dominates the laureates of these prizes.

However, the Nobel Prize does not give awards in computer science, which raise the question as to the common characteristics that distinguished computer scientists share. Unlike the Nobel Prize, where the subjects were well-established, the Turing Award celebrates the influential scientists in computer science, which did not become a distinct academic discipline until 70 years ago \cite{CS}. By studying the key characteristics of the Turing Award winners, we will shed light on the ways to become top scientists in the prospering area of computer science. Like the Nobel Prize laureates, the winners of the Turing Award are well acknowledged high achievers in the field of computer science. These laureates’ achievements and reasons for winning the award are clearly stated, and by the nature of the award, the sample will be spread over time and regions. Therefore, the Turing Award laureates are good representation of the greatest computer scientists in the world since the establishment of the subject, and the award creates a good sample for studying the common characteristics of distinguished computer scientists. 

Compared to Nobel Prize, the work focused on the Turing Award is limited. From a sociology study, \citet{akmut} concluded that place of birth, nationality, gender, social background, race and networks play a role in making Turing Award laureates. \citet{xia2020turing} proposed a metric named "Turing Number" to measure the distance between scholars and Turing Award laureates, as a new way to construct a scientific collaboration network centered around the laureates. The Turing Award laureates also provide good performance when being used as a data source for studying and testing other models and metrics. To understand the process of scientific innovation, \citet{liu} studied 70 Turing Award winners and developed a chaining model that showed the academic career path of these computer scientists. In order to compare PageRank-based and citations-based rankings of scientists, \citet{pagerank} used the Codd Award and Turing Award winners and found that citations-based rankings perform better for Codd Award winners, but PageRank-based methods do so for Turing Award recipients when using absolute ranks.
Not only the characteristics of Turing Award laureates are worth studying, the award also provides a good data source for testing and evaluating other algorithms if needed. The current research regarding the Turing Award are mostly focused on some of its specific aspects, yet a thorough analysis of the laureates from the aspect of data science is needed. Therefore, we hope to construct a comprehensive dataset and analyze the frequent features to pave the way for future studies regarding top computer scientists. 

Since citations is one of the common ways to evaluate scientists' academic accomplishments, one goal of our study is to compare different sources of citations on their ability to predict the winners. It is observed that the main citation sources in computer science are Google Scholar, Semantic Scholar and the ACM Author Profile. However, these three sources exhibit significant differences in the citation counts. The ACM Author Profile has the least average citation count among the three  sources. Google Scholar is found to include patents and even non-academic documents, while Semantic Scholar indexes academic papers but not other content such as patents \cite{citation}. A recent study \cite{compare} proved that the artificial intelligence engine used in Semantic Scholar enables it to understand the context of the document more deeply and thus provides an accurate search result while the results are limited and not up-to-date. By collecting the citation data from the Turing Award laureates, we hope to shed light on how different sources of citations reflect the computer scientists' accomplishments. The author profiles of 41 Turing Award laureates can not be found on Google Scholar, while no one is missing from Semantic Scholar and ACM Author Profile.

\section{Methods}
In this section, we will first describe how we collect information of each of the Turing Award laureates and how we construct the dataset for our study. We then explain the technologies we use in data analysis. 

\subsection{Data Preparation}
To construct a dataset that reflects the Turing Award laureates' experiences and backgrounds, we select features in the following four aspects: personal information, family background, academic background, and industry experiences. 

Personal information includes: birth year, gender, race, and citizenship. Race is denoted by White, Jewish and Asian. Laureates with dual citizenship are counted as citizens for both of the countries. 

Family background includes: number of children, marital status, parents' background, number of siblings, and order of birth. Marital status are specified as number of marriages if applicable. Parents' backgrounds are either a short paragraph or several sentences about the laureates' family. Order of birth are specified to younger (if one sibling), elder (if one sibling), youngest (if more than one sibling), middle (if more than one sibling), and eldest (if more than one sibling). 

Academic background includes: highest degree, universities attended, majors studied, location of education, fields of study, citation count, PhD advisor, and university affiliation. Since honorary degrees are usually given when the recipients already have distinguished achievements, honorary degrees are not included as part of the educational background. Majors studied include undergraduate majors and graduate majors. Same majors are counted for undergraduate and graduate, respectively, while same majors for more than one graduate programs are only counted once. Double majors are included as well. To reduce the range of our data and focus on the general major field, the majors "mathematics", "applied mathematics" and "numerical mathematics" are grouped together as "mathematics". Also, majors such as "computer science and electrical engineering" are separated into "computer science" and "electrical engineering". Location of education is specified as countries, with no distinction between undergraduate study and graduate study locations. Fields of study are the fields in computer science that the laureates have made significant contributions to and are mentioned in the rationale. University affiliation includes both long-term affiliation as well as short-term affiliation. 

Industry experiences include whether they have had industry experience and their affiliations at the time of award. Affiliation at the time of award is later binned into four categories: educational institution, industry, both educational institution and industry, and government. 

The main source of our data is the ACM official website of the Turing Award \cite{TuringOfficial}. On the main page of most of the laureates, their birthday, educational path, past and current affiliations, as well as a short biography are shown. Missing information are manually collected from the Wikipedia page of the laureates. 

The laureates' academic background are collected from Google Scholar \cite{GoogleScholar}, Semantic Scholar \cite{SemanticScholar} and ACM Author Profile \cite{ACM}. A laureate's citation count for the most cited paper, the top five cited papers and total citations are collected from each of the three sources. While the profile for each laureate can be found on Semantic Scholar and ACM Author Profile, only 43 percent of the laureates are documented by Google Scholar. The family backgrounds are first collected from the biographies on the laureates' ACM main pages and then supplemented by their interview transcripts if available. 74 percent of the laureates' description of their family backgrounds are found, while only half of them mention their birth orders. All of the laureate's marital status are collected. Overall, despite the missing laureates in Google Scholar and the missing values in family background, all of the other features are successfully collected for all the 72 laureates. 

\subsection{Data Analysis}
To see the general features of the Turing Award laureates, we implement the frequent pattern growth (FP-Growth) algorithm \cite{fpgrowth} for pattern mining. It adopts a divide-and-conquer strategy to mine the frequent itemsets without costly candidate generation. After a frequent pattern tree is generated from the whole dataset, it then recursively generates and mines each of the conditional frequent pattern trees for frequent patterns, which significantly reduces the time complexity of pattern mining \cite{textbook}. In our study, we group the features for each laureate as an itemset and utilize the FP-Growth algorithm to extract the frequent patterns of these features. To avoid low support, we select features that have small variations, including sex, citizenship, race, marital status, order of birth, highest education level, schools attended, field of study, university affiliation, majors, education location, and whether they have industry experience and affiliation at the time of award (grouped). Using this subset of our dataset, the frequent patterns of features of the laureates are generated. 

The features are then separately analyzed. Logistic regression plot, pie chart, word cloud and map are generated accordingly for each of the interesting features we find.

\section{Results}
In this section, we show our findings by analyzing Personal Information, Family Background, Academic Background and Industry Experience of the Turing Award Laureates.

\subsection{Turing Award General Information}
\begin{figure}[htbp]
    \centering
    \subfigure[]{
    \label{fig0:sub:1}
    \includegraphics[width=0.98\columnwidth]{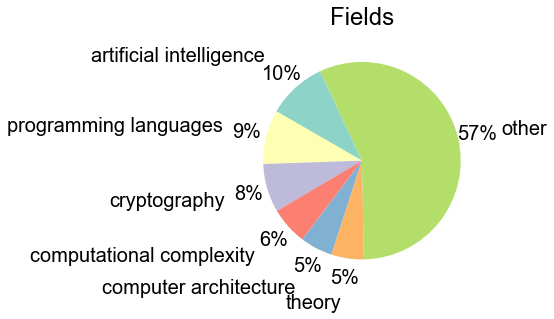}}
    \subfigure[]{
    \label{fig0:sub:2}
    \includegraphics[width=1.0\columnwidth]{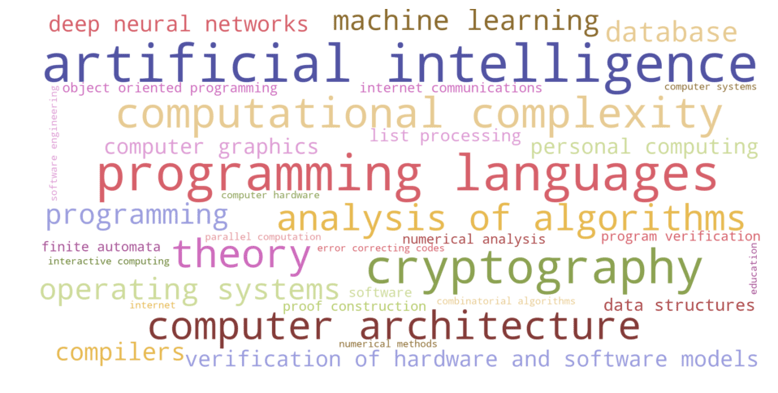}}
    \caption{Turing Awards by fields.}
    \label{fig:0}
\end{figure}

The Turing Award has been awarded 54 times with 72 winners as of 2020. Among them, 16 times were team awards (13 times with two winners, and 3 times with three winners). Solo award is the most common, accounting for 72\% of the total number of awards.
Fig. \ref{fig:0} shows their awarded fields, where 31 people were awarded in more than one field. Artificial Intelligence, Programming Languages, and Cryptography are the most popular fields being awarded, followed by Computational Complexity and Theory. A total of 45 scientists in the above five fields have won the Turing Award.

\subsection{Personal Information}
\begin{figure}[htbp]
    \centering
    \includegraphics[width=1.0\columnwidth]{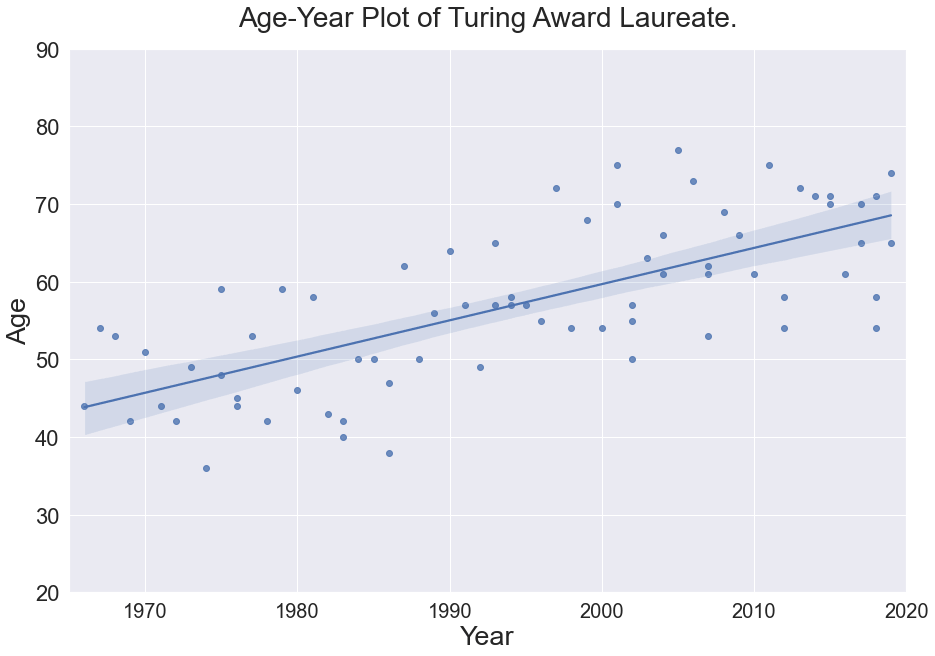}
    \caption{Age-Year relationship of Turing Award laureates. The blue line is fitted by Logistic Regression, and the shadow part shows the 95\% confidence interval for the regression.}
    \label{fig:1}
\end{figure}

The relationship between the age of the laureates and the year of the award are shown in Fig. \ref{fig:1}. Logistic regression is used to fit the linear relationship between age and year of the award by minimizing the Mean Squared Error (MSE). The age and year of award are positively correlated, i.e., the age at which a laureate won the award increases over time. The average age of the scientist at the time of the award is 57.25. Donald Knuth \cite{Donald} is the youngest and Peter Naur \cite{Peter} is the eldest when they were awarded, at the age of 36 and 77, respectively.

\begin{figure}[htbp]
    \centering
    \includegraphics[width=1.0\columnwidth]{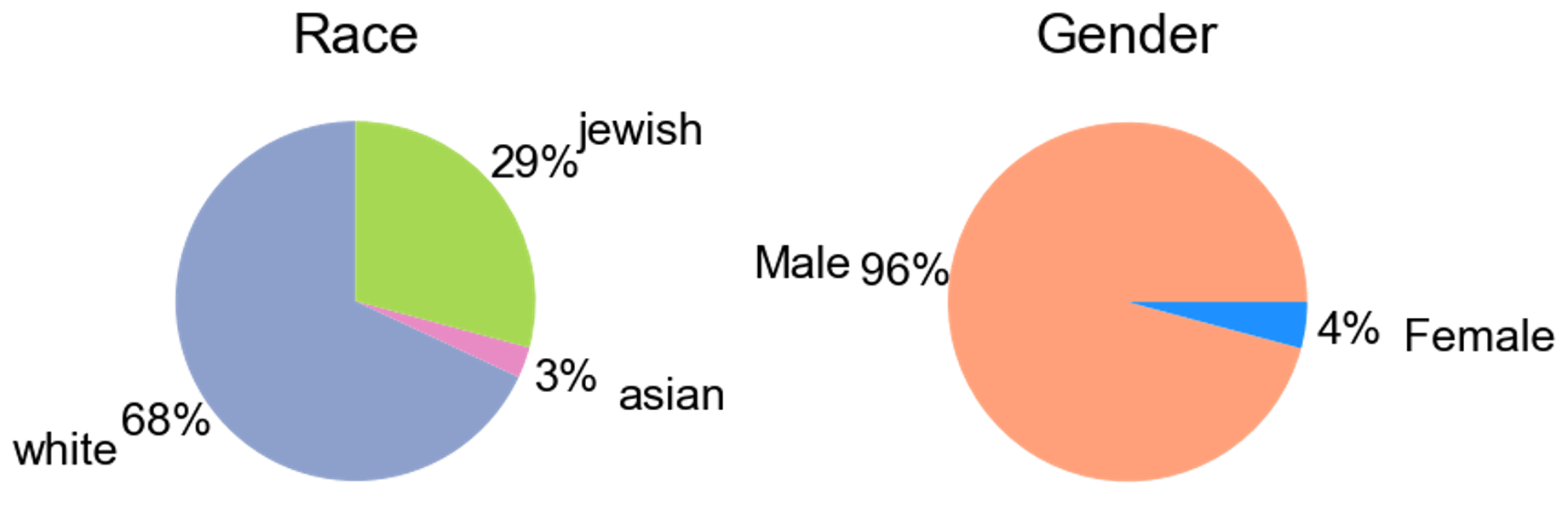}
    \caption{Race and gender of the Turing Award laureates.}
    \label{fig:2}
\end{figure}

In terms of race and gender, as shown in Fig. \ref{fig:2}, white males dominate the Turing Award. Among the 72 winners, 49 of them are white, and 21 of them are Jewish; only three females have won the award. The first woman who won the Turing Award was Frances Elisabeth Allen, in 2006, 41 years after the award was established.

\begin{figure}[htbp]
    \centering
    \includegraphics[width=1.0\columnwidth]{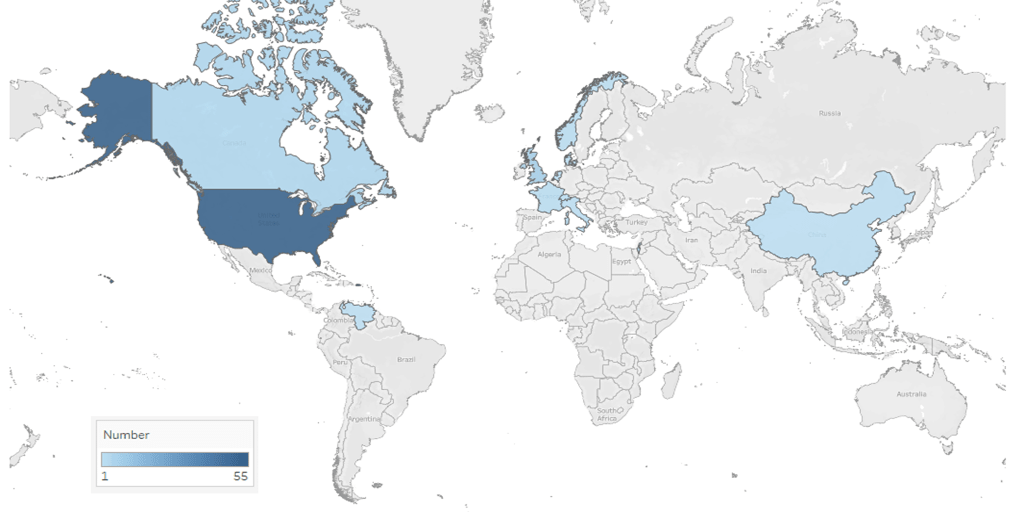}
    \caption{Distribution of the Turing Award laureates' citizenship.}
    \label{fig:3}
\end{figure}

We show the citizenship of the Turing Award laureates in Fig. \ref{fig:3}. Clearly, the United States is the country that is the most advanced in computer science. Among the 72 Turing Award winners, 55 are United States citizens and 8 are citizens of the United Kingdom. Followed by Israel and Canada with with 5 winners each.

\subsection{Family Background}
\begin{figure}[htbp]
    \centering
    \includegraphics[width=1.0\columnwidth]{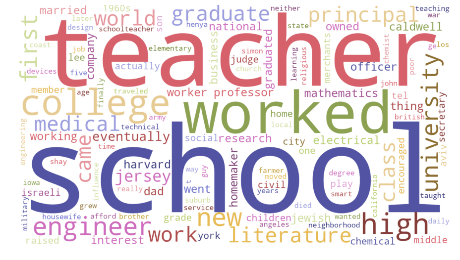}
    \caption{Word cloud when the Turing Award laureates talk about their family.}
    \label{fig:5}
\end{figure}

We extract the winner's family information mentioned in the interview transcript and generate a word cloud based on the word frequency, as shown in Fig. \ref{fig:5}. The most common words are "teacher", "school", "worked" and "college", which indicate that the Turing award laureates are likely to have good family conditions, and that their parents' professions are mostly related to education. All winners are heterosexual and have been married at least once. 

\begin{figure}[htbp]
    \centering
    \includegraphics[width=1.0\columnwidth]{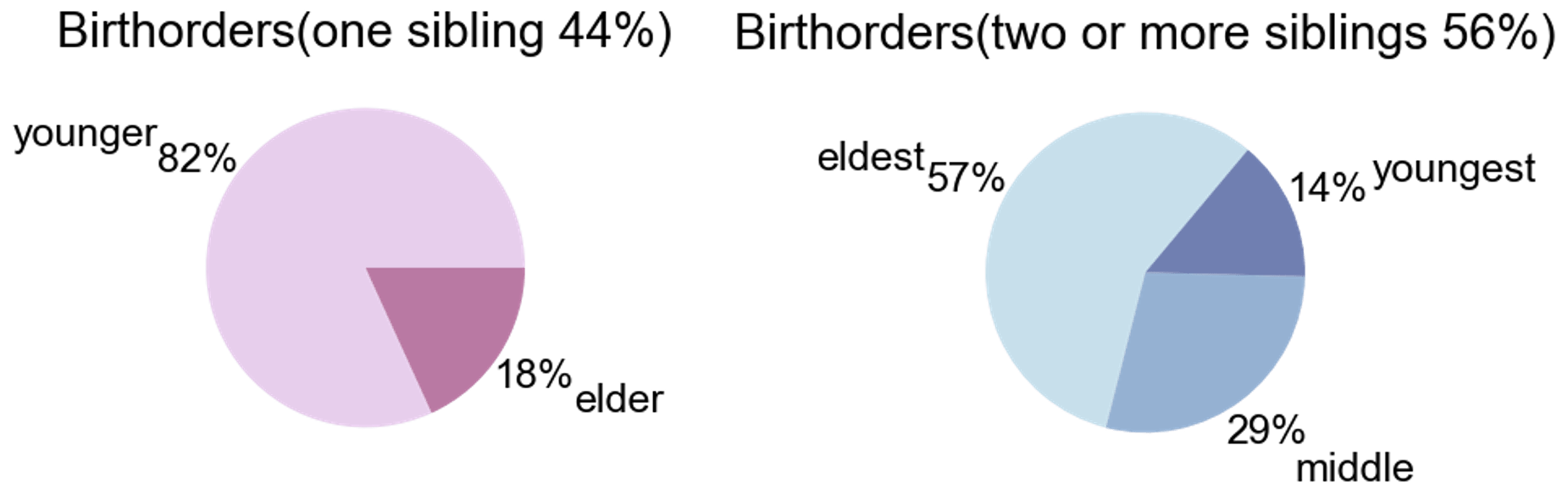}
    \caption{Birth order of the Turing Award laureates.}
    \label{fig:4}
\end{figure}

The relationship between birth order and personal achievement has always been a controversial topic. \citet{birth1929} show that first-born and second-born children have IQ differences, and \citet{clark} show that the birth order has impact on the Nobel Prize winners. The order of birth information of the Turing award laureates is shown in Fig. \ref{fig:4}. We find that if the winner only has one sibling, then he/she is much more likely to be the younger one; if the winner has more than one sibling, then he/she is the most likely to be the eldest one. Our findings are consistent with the research by \citet{clark}.

\begin{figure}[htbp]
    \centering
    \includegraphics[width=1.0\columnwidth]{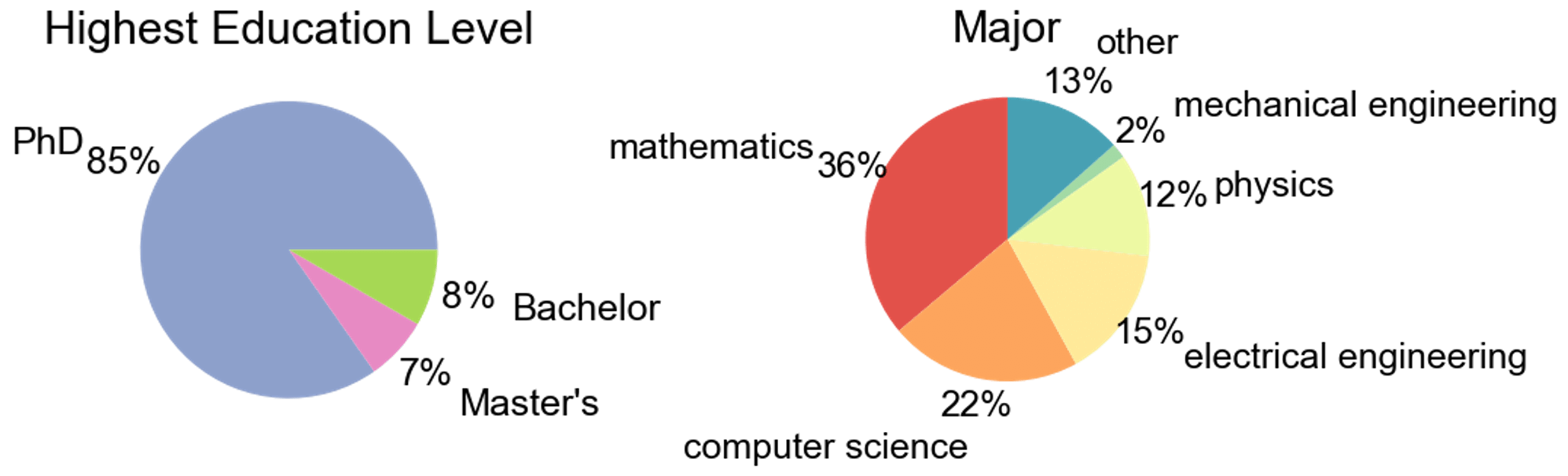}
    \caption{Educational background.}
    \label{fig:6}
\end{figure}

\subsection{Academic Background}

\begin{figure}[htbp]
    \centering
    \subfigure[]{
    \label{fig7:sub:1}
    \includegraphics[width=0.85\columnwidth]{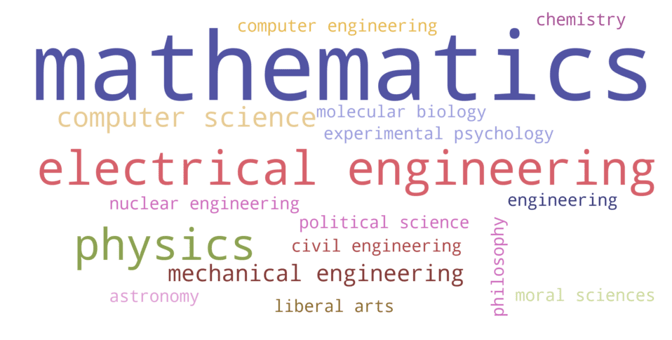}}
    \subfigure[]{
    \label{fig7:sub:2}
    \includegraphics[width=0.85\columnwidth]{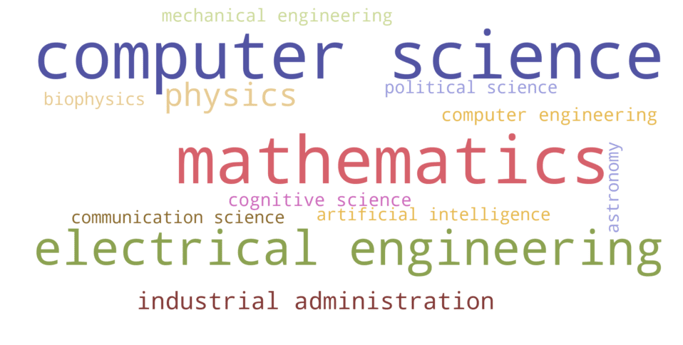}}
    \caption{Word cloud showing the distribution of the majors of the Turing Award laureates: (a) undergraduate majors, (b) graduate majors.}
    \label{fig:7}
\end{figure}

Fig. \ref{fig:6} shows the highest degree the winners obtained and their majors. Among all the 72 Turing Award winners, 61 obtained a PhD degree, 5 obtained a master's degree and 6 obtained a bachelor's degree. Regarding their majors, about one third of the awarded scientists studied mathematics during their higher education. We continue to show detail about the winners' undergraduate major and graduate major, as shown in Fig. \ref{fig:7}. Surprisingly, only three winners had a major in Computer Science when they were undergraduate, and they were all double majors. This could be explained by the fact that computer science is a relatively new major. The world's first computer science degree program, the Cambridge Diploma in Computer Science, began at the University of Cambridge Computer Laboratory in 1953. The first computer science department in the United States was formed at Purdue University in 1962. More than half of the scientists studied mathematics during their undergraduate studies. In their graduate studies, computer science became the most popular major, while many people continued to study mathematics.

\begin{figure}[htbp]
    \centering
    \includegraphics[width=0.5\textwidth]{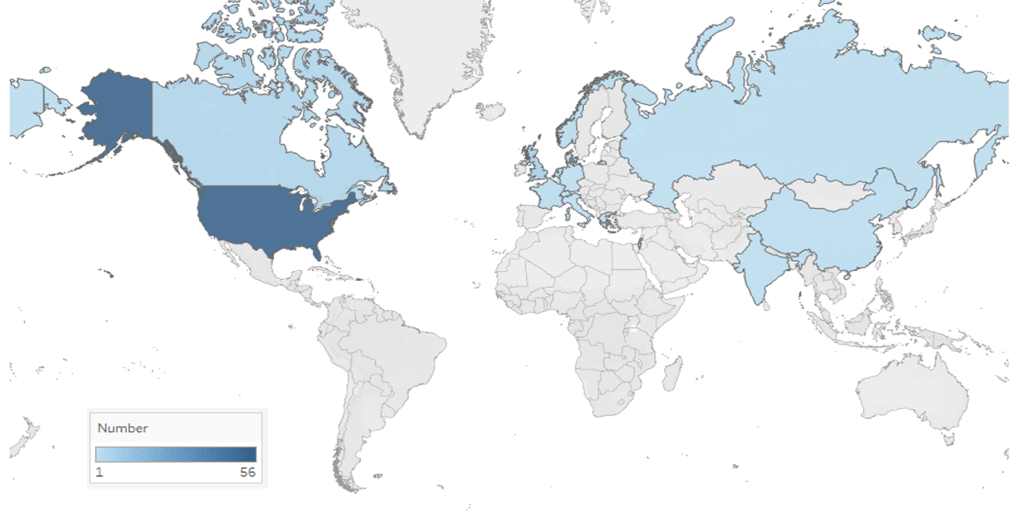}
    \caption{Distribution of the Turing Award laureates' education locations. Darker colors indicate more people. }
    \label{fig:8}
\end{figure}

We further show where they received their degrees in Fig. \ref{fig:8}, where the darker the color is, the more people there are. Clearly, the United States remains the most dominant country. We also show the top 12 universities where the Turing Award winners graduated in Table \ref{tab:1}, sorted by number. University of California at Berkeley has the most number of Turing Award winners among all schools in both undergraduate and graduate programs. 
Following that, the most popular universities that the winners did their undergraduate studies include: University of Cambridge, Harvard University and Carnegie Mellon University. The most popular universities that the winners did their graduate studies include: University of California at Berkeley, Harvard University, Princeton University, and Stanford University. Stanford University has attracted the most number of Turing Award winners to teach there, with the number of 22, although only two undergraduate and six graduate students are from Stanford University .

\begin{figure*}[htbp]
    \centering
    \subfigure[]{
    \label{figc:sub:1}
    \includegraphics[width=0.9\textwidth]{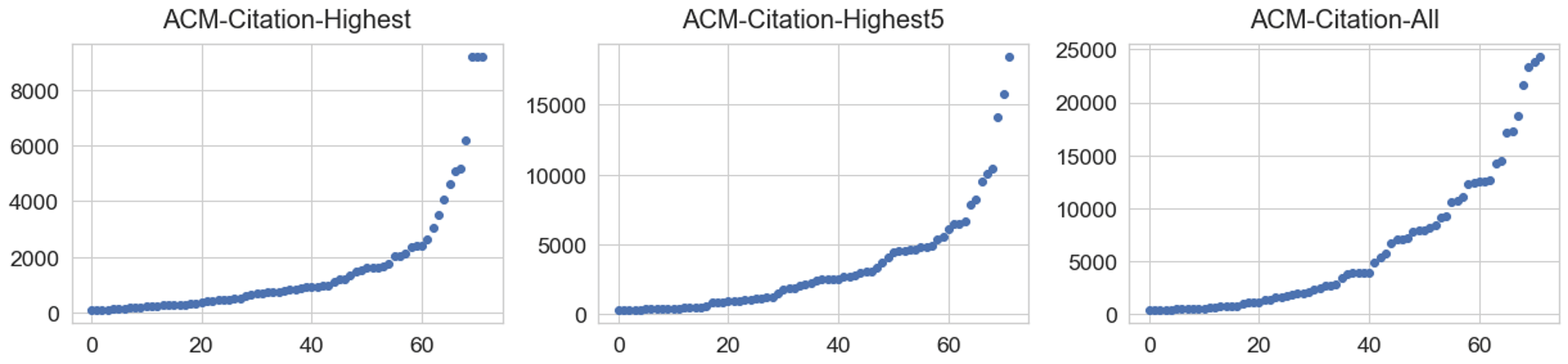}}
    \subfigure[]{
    \label{figc:sub:2}
    \includegraphics[width=0.9\textwidth]{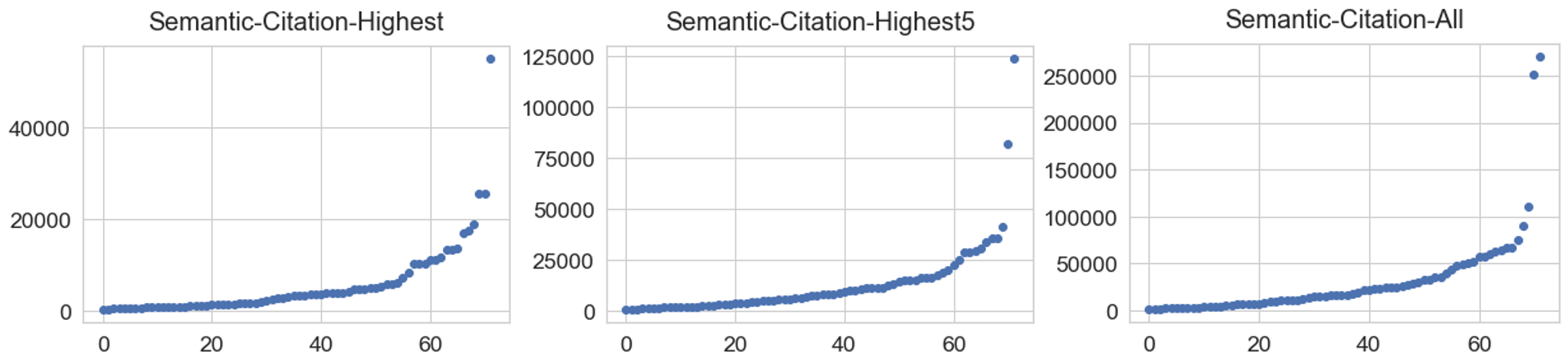}}
    \subfigure[]{
    \label{figc:sub:3}
    \includegraphics[width=0.9\textwidth]{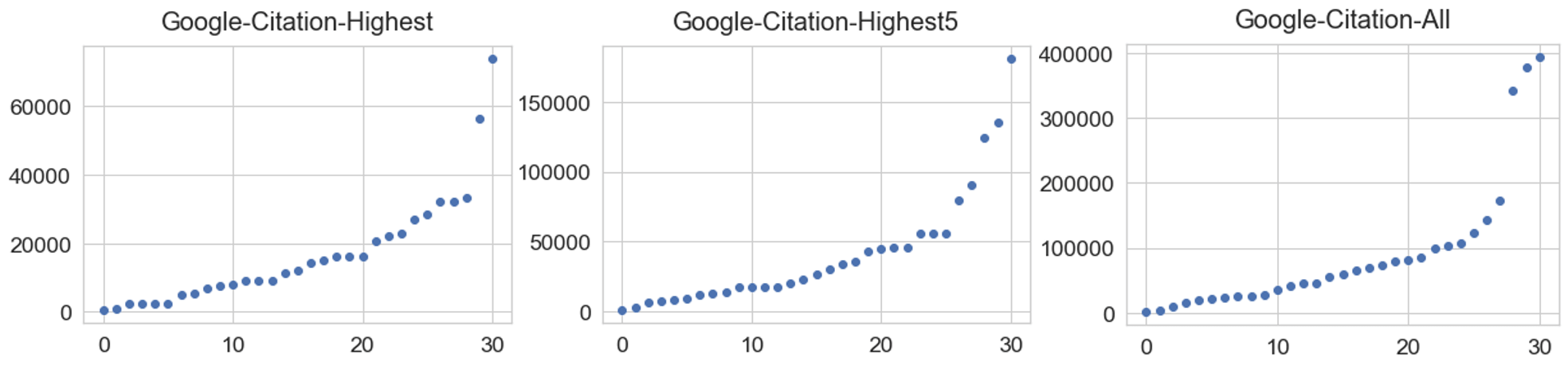}}
    \caption{Sorted citations from ACM Author Profile, Semantic Scholar, and Google Scholar: (a) ACM Author Profile, (b)Semantic Scholar, and (c) Google Scholar.}
    \label{fig:c}
\end{figure*}

Citation is often a way to evaluate a scholar's academic influence. We collect the winners citation information, including the most cited paper, the sum of top 5 paper citations, and the total citations, as of November 30, 2020. Since different databases have different standards for citation, we choose to use Semantic Scholar, Google Scholar and ACM Author Profile as our resources. In order to make the result more intuitive, we sort the citations and the results are shown in Fig. \ref{fig:c}. All the citation measures show a similar exponential distribution, regardless the resource. Only a few of them have a very high citation count and these people are all in the Artificial Intelligence area. We compare the rankings of the Turing Award winners among all computer science scholars given by \cite{guide2research}, and only five of the Turing Award Laureates are listed in top 100 computer scientists. The results indicate that one may not need a very high citation to win the award, for example, Tim Berners-Lee, the inventor of the World Wide Web, won the Turing Award in 2016 with under 2,000 citations.

It is worth noting that 19 of the 72 laureates have an advisor-student relationship as shown in Fig. \ref{fig:10}.

\begin{figure}[htbp]
    \centering
    \includegraphics[width=0.96\columnwidth]{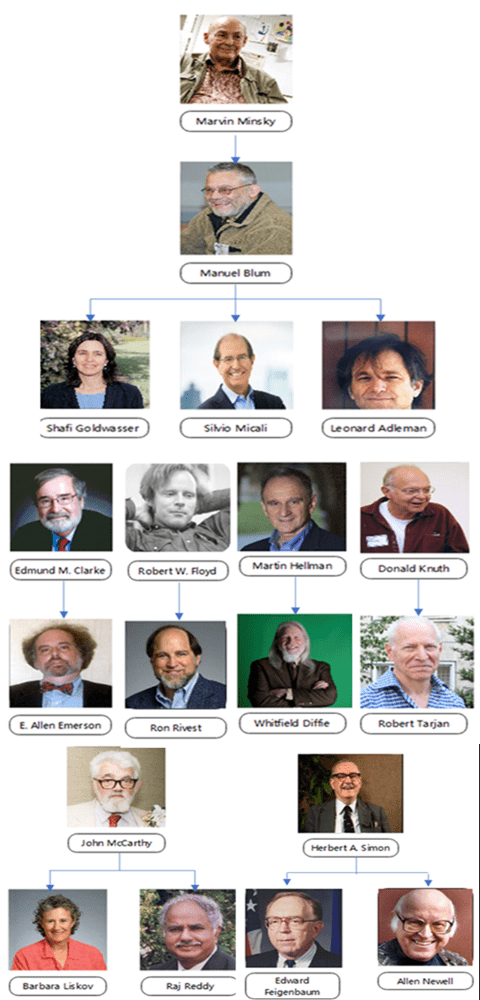}
    \caption{Advisor-student relationship.}
    \label{fig:9}
\end{figure}

\subsection{Industry Experience}

% \begin{figure}[htbp]
%     \centering
%     \subfigure[]{
%     \label{fig10:sub:1}
%     \includegraphics[width=0.16\textwidth]{indutry.png}}
%     \subfigure[]{
%     \label{fig10:sub:2}
%     \includegraphics[width=0.28\textwidth]{affiliation.png}}
%     \caption{Industry Experiences of Turing Award Laureates.}
%     \label{fig:10}
% \end{figure}

\begin{figure}[htbp]
    \centering
    \includegraphics[width=1.0\columnwidth]{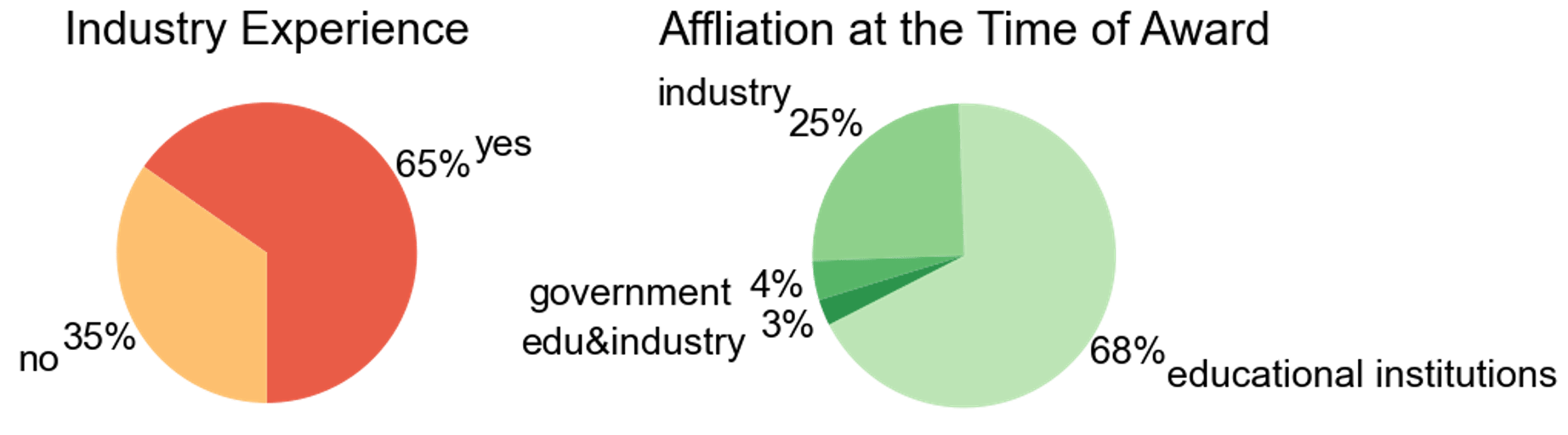}
    \caption{Industry experiences of the Turing Award laureates.}
    \label{fig:10}
\end{figure}

As shown in Fig. \ref{fig:10}, about 65\% of the scientists had industry experience, but most of them eventually went to academia. At the time they received the Turing Award, about 68\% of them were working in universities.

\section{Conclusion and Future Work}
In this study, we build a comprehensive dataset of the Turing Award laureates, including their personal information, family background, academic background, and industry experience. We then investigate  potential patterns in the personal background of these Turing Award winners. Our analysis results show that most scientists who won the Turing Award have the following characteristics: white, male, married, United States citizen, and having a PhD degree. Four particular findings are interesting: the age at which a laureate won the award increases over time, most of the Turing Award laureates did not major in computer science even though the award celebrates the highest achievements in computer science, the order of birth has strong correlations with the winners' success, and the number of citations is not as important as one would expect.

We can consider three directions in our future work: 1) collecting more comprehensive data of the laureates' social networks, including their collaborators, and apply graph mining techniques \cite{chakrabarti2006graph} to mine the winners' social networks, 2) counting the average citations in different areas of computer science and normalize the citation data for each area, and 3) including top computer scientists who have not won the Turing Award in our dataset and utilizing machine learning algorithms to predict future Turing Award winners.

\begin{table*}
    \centering
    \caption{The Number of Turing Award Laureates by universities.}  
    \scalebox{0.79}{
    \begin{tabular}{|c|c||c|c||c|c|}
    \hline
        Undergraduate Universities &  Count & Graduate Universities & Count & Working Universities & Count\\
    \hline
    \hline
        University of California, Berkley & 6 & University of California, Berkley & 7 & Stanford University & 22 \\
        University of Cambridge & 5 & Princeton University & 7 & Massachusetts Institute of Technology & 19 \\
        Harvard University & 4 & Harvard University & 7 & University of California, Berkley & 13 \\
        Carnegie Mellon University & 4 & Stanford University & 6 & Carnegie Mellon University & 10 \\
        Massachusetts Institute of Technology & 3 & Massachusetts Institute of Technology & 4 & Princeton University & 7 \\
        California Institute of Technology & 3 & Weizmann Institute of Science & 2 & Harvard University & 7 \\
        University of Oxford & 3 & University of Michigan & 2 & New York University & 5 \\
        Duke University & 3 & California Institute of Technology & 2 & University of Oxford & 4 \\
        University of Chicago & 3 & University of Oslo & 2 & University of Toronto & 4 \\
        Technion–Israel Institute of Technology & 2 & University of Toronto & 2 & Weizmann Institute of Science & 3 \\
        Stanford University & 2 & University of Illinois at Urbana–Champaign & 2 & Cornell University & 3 \\
        University of Oslo & 2 & Carnegie Mellon University & 2 & University of Edinburgh & 2 \\
    \hline
    \end{tabular}}

    \label{tab:1}
\end{table*}

%\section{Acknowledgments} \\cannot have this due to double blind review
%The authors acknowledge the help of professor Jiebo Luo, teaching assistants Hanjia Lyu and Songyang Zhang at the University of Rochester.

% \bibliographystyle{aaai}
\bibliography{reference}

\begin{thebibliography}{30}
\providecommand{\natexlab}[1]{#1}
\providecommand{\url}[1]{\texttt{#1}}
\providecommand{\urlprefix}{URL }
\expandafter\ifx\csname urlstyle\endcsname\relax
  \providecommand{\doi}[1]{doi:\discretionary{}{}{}#1}\else
  \providecommand{\doi}{doi:\discretionary{}{}{}\begingroup
  \urlstyle{rm}\Url}\fi

\bibitem[{ACM(2020{\natexlab{a}})}]{ACM}
ACM. 2020{\natexlab{a}}.
\newblock ACM Author Profile Page.
\newblock [Online, accessed 3-November-2020]. Available:
  \url{https://www.acm.org/publications/acm-author-profile-page}.

\bibitem[{ACM(2020{\natexlab{b}})}]{TuringOfficial}
ACM. 2020{\natexlab{b}}.
\newblock ALPHABETICAL LISTING OF A.M. TURING AWARD WINNERS.
\newblock [Online, accessed 1-November-2020]. Available:
  \url{https://amturing.acm.org/alphabetical.cfm}.

\bibitem[{ACM(2020{\natexlab{c}})}]{TuringNominate}
ACM. 2020{\natexlab{c}}.
\newblock HOW TO NOMINATE.
\newblock [Online, accessed 1-November-2020]. Available:
  \url{https://amturing.acm.org/call_for_nominations.cfm}.

\bibitem[{ACM(2020{\natexlab{d}})}]{guide2research}
ACM. 2020{\natexlab{d}}.
\newblock Ranking for Computer Science.
\newblock [Online, accessed 9-Dec-2020]. Available:
  \url{http://www.guide2research.com/scientists/}.

\bibitem[{Akmut(2018)}]{akmut}
Akmut, C. 2018.
\newblock Social conditions of outstanding contributions to computer science: a
  prosopography of Turing Award laureates (1966-2016) .

\bibitem[{Arum(2016)}]{compare}
Arum, N.~S. 2016.
\newblock A look at semantic scholar and Google scholar.

\bibitem[{Berry(1981)}]{berry}
Berry, C. 1981.
\newblock The Nobel scientists and the origins of scientific achievement.
\newblock \emph{The British Journal of Sociology} 32(3): 381--391.

\bibitem[{Chakrabarti and Faloutsos(2006)}]{chakrabarti2006graph}
Chakrabarti, D.; and Faloutsos, C. 2006.
\newblock Graph mining: Laws, generators, and algorithms.
\newblock \emph{ACM computing surveys (CSUR)} 38(1): 2--es.

\bibitem[{Charlton(2007)}]{charlton}
Charlton, B.~G. 2007.
\newblock Which are the best nations and institutions for revolutionary science
  1987--2006? Analysis using a combined metric of Nobel prizes, Fields medals,
  Lasker awards and Turing awards (NFLT metric).

\bibitem[{Clark and Rice(1982)}]{clark}
Clark, R.~D.; and Rice, G.~A. 1982.
\newblock Family constellations and eminence: The birth orders of Nobel Prize
  winners.
\newblock \emph{The Journal of Psychology} 110(2): 281--287.

\bibitem[{Clynes(2016)}]{nobel2016}
Clynes, T. 2016.
\newblock Where Nobel winners get their start.
\newblock \emph{Nature News} 538(7624): 152.

\bibitem[{Cooper and Van~Leeuwen(2013)}]{alanturing}
Cooper, S.~B.; and Van~Leeuwen, J. 2013.
\newblock \emph{Alan Turing: His work and impact}.
\newblock Elsevier.

\bibitem[{Fiala and Tutoky(2017)}]{pagerank}
Fiala, D.; and Tutoky, G. 2017.
\newblock PageRank-based prediction of award-winning researchers and the impact
  of citations.
\newblock \emph{Journal of Informetrics} 11(4): 1044--1068.

\bibitem[{Goldstein and Brown(2012)}]{golden}
Goldstein, J.~L.; and Brown, M.~S. 2012.
\newblock A golden era of Nobel laureates.
\newblock \emph{Science} 338(6110): 1033--1034.

\bibitem[{Google(2020)}]{GoogleScholar}
Google. 2020.
\newblock Google Scholar.
\newblock [Online, accessed 2-November-2020]. Available:
  \url{https://scholar.google.com}.

\bibitem[{Han, Pei, and Kamber(2011)}]{textbook}
Han, J.; Pei, J.; and Kamber, M. 2011.
\newblock \emph{Data mining: concepts and techniques}.
\newblock Elsevier.

\bibitem[{Han, Pei, and Yin(2000)}]{fpgrowth}
Han, J.; Pei, J.; and Yin, Y. 2000.
\newblock Mining frequent patterns without candidate generation.
\newblock \emph{ACM sigmod record} 29(2): 1--12.

\bibitem[{Liu and Xu(2020)}]{liu}
Liu, E.; and Xu, Y. 2020.
\newblock Chaining and the process of scientific innovation .

\bibitem[{Rothenberg(2005)}]{family}
Rothenberg, A. 2005.
\newblock Family background and genius II: Nobel laureates in science.
\newblock \emph{The Canadian Journal of Psychiatry} 50(14): 918--925.

\bibitem[{Schlagberger, Bornmann, and Bauer(2016)}]{schlag}
Schlagberger, E.~M.; Bornmann, L.; and Bauer, J. 2016.
\newblock At what institutions did Nobel laureates do their prize-winning work?
  An analysis of biographical information on Nobel laureates from 1994 to 2014.
\newblock \emph{Scientometrics} 109(2): 723--767.

\bibitem[{SemanticScholar(2020{\natexlab{a}})}]{citation}
SemanticScholar. 2020{\natexlab{a}}.
\newblock How does Semantic Scholar calculate citation counts?
\newblock [Online, accessed 10-Novemeber-2020]. Available:
  \url{https://www.semanticscholar.org/faq#estimated-citations}.

\bibitem[{SemanticScholar(2020{\natexlab{b}})}]{SemanticScholar}
SemanticScholar. 2020{\natexlab{b}}.
\newblock Semantic Scholar.
\newblock [Online, accessed 4-November-2020]. Available:
  \url{https://www.semanticscholar.org}.

\bibitem[{t2informatic(2020)}]{t2informatic}
t2informatic. 2020.
\newblock A.M. Turing Award.
\newblock [Online, accessed 9-Dec-2020]. Available:
  \url{https://t2informatik.de/en/smartpedia/turing-award/}.

\bibitem[{Thurstone and Jenkins(1929)}]{birth1929}
Thurstone, L.~L.; and Jenkins, R.~L. 1929.
\newblock Birth order and intelligence.
\newblock \emph{Journal of Educational Psychology} 20(9): 641.

\bibitem[{Wagner et~al.(2015)Wagner, Horlings, Whetsell, Mattsson, and
  Nordqvist}]{wagner}
Wagner, C.~S.; Horlings, E.; Whetsell, T.~A.; Mattsson, P.; and Nordqvist, K.
  2015.
\newblock Do Nobel Laureates create prize-winning networks? An analysis of
  collaborative research in physiology or medicine.
\newblock \emph{PloS one} 10(7): e0134164.

\bibitem[{Wikipedia(2020{\natexlab{a}})}]{CS}
Wikipedia. 2020{\natexlab{a}}.
\newblock Computer Science.
\newblock [Online, accessed 10-December-2020]. Available:
  \url{https://en.wikipedia.org/wiki/Computer_science}.

\bibitem[{Wikipedia(2020{\natexlab{b}})}]{digitalrev}
Wikipedia. 2020{\natexlab{b}}.
\newblock Digital Revolution.
\newblock [Online, accessed 9-Dec-2020]. Available:
  \url{https://en.wikipedia.org/wiki/Digital_Revolution}.

\bibitem[{Wikipedia(2020{\natexlab{c}})}]{Donald}
Wikipedia. 2020{\natexlab{c}}.
\newblock Donald Knuth.
\newblock [Online, accessed 2-Dec-2020]. Available:
  \url{https://en.wikipedia.org/wiki/Donald_Knuth}.

\bibitem[{Wikipedia(2020{\natexlab{d}})}]{Peter}
Wikipedia. 2020{\natexlab{d}}.
\newblock Peter Naur.
\newblock [Online, accessed 2-Dec-2020]. Available:
  \url{https://en.wikipedia.org/wiki/Peter_Naur}.

\bibitem[{Xia et~al.(2020)Xia, Liu, Ren, Wang, and Kong}]{xia2020turing}
Xia, F.; Liu, J.; Ren, J.; Wang, W.; and Kong, X. 2020.
\newblock Turing number: how far are you to AM Turing award?
\newblock \emph{ACM SIGWEB Newsletter} (Autumn): 1--8.

\end{thebibliography}

\end{document}